\documentclass[preprint,showpacs,preprintnumbers,amsmath,amssymb,floatfix,prc]{revtex4}
                                                                                
\usepackage{graphicx}
\usepackage{dcolumn}
\usepackage{bm}
                                                                                
\begin{document}

\title{$J/\psi$ production from charm coalescence in relativistic heavy ion collisions}

\author{Bin Zhang}

\affiliation{Department of Chemistry and Physics, 
Arkansas State University,
State University, AR 72467-0419, USA}

\date{June 20, 2006}

\begin{abstract}
$J/\psi$ production and collective flow is studied with a coalescence 
model based on phase space distribution of charm quarks from a 
multi-phase transport model simulation of relativistic heavy ion 
collisions. Both the yield and the flow of $J/\psi$ particles are 
sensitive to charm quark final state interactions. 
As the charm quark rescattering cross section increases from 3 mb to
10 mb, $J/\psi$ elliptic flow increases faster than corresponding
light hadron elliptic flows. The $v_2(p_t)$ of $J/\psi$ crosses that
of $D$ mesons to reach a value that is about the peak value
of the $D$ meson flow but at a higher $p_t$. 
As $J/\psi$ elliptic flow has only contributions from charm
quarks, it complements $D$ meson elliptic flow in reflecting charm
properties in the Quark-Gluon Plasma.
\end{abstract}


\maketitle


Experimental data from the Relativistic Heavy Ion Collider (RHIC)
at Brookhaven National Lab (BNL) revealed fascinating properties
of the Quark-Gluon Plasma (QGP) produced in relativistic nucleus-nucleus
collisions \cite{Arsene:2004fa,Back:2004je,Adams:2005dq,%
Adcox:2004mh,Gyulassy:2004zy,Shuryak:2004cy}. 
The $J/\psi$ particle is one of the important probes
of QGP properties. It was proposed as a signature of QGP formation
because of its dissociation due to color screening inside the
Quark-Gluon Plasma \cite{Matsui:1986dk}. 
Dissociations due to comover scatterings
have also been studied to interpret experimental
data at SPS energies \cite{Gavin:1996yd,Kharzeev:1996yx,%
Capella:2000zp}. Recently, 
lattice Quantum Chromodynamics (QCD) 
calculations show that the $J/\psi$ particle can survive the
plasma up to about $2T_c$ \cite{Asakawa:2003re,Datta:2003ww}. 
The survival of $J/\psi$ was also shown by potential 
models \cite{Wong:2004zr}. This leads to new insights into
the experimental data \cite{Karsch:2005nk}. 
At RHIC energies, many pairs of 
charm and anti-charm quarks can be produced in a single event.
These charm and anti-charm quarks may recombine into $J/\psi$ particles.
The recombination can contribute significantly to the final
$J/\psi$ yield \cite{Thews:2000rj,Thews:2005vj,%
Braun-Munzinger:2000px,Andronic:2003zv,%
Grandchamp:2001pf,Grandchamp:2002wp,Grandchamp:2003uw,%
Zhang:2002ug,Zhang:2003dp,%
Bratkovskaya:2003ux,Greco:2003vf}. 
In this paper, $J/\psi$ production and flow
will be studied by a phase-space coalescence model using
the charm freeze-out information from a multi-phase transport (AMPT)
model. In the following, the AMPT model and 
the coalescence formalism will be reviewed. 
This is followed by the 
presentation of results including the $J/\psi$ yield,
$\langle p_t^2 \rangle$, $p_t$ distributions, and $v_2(p_t)$.
Finally, a summary will be given with the implications
of these results on the QGP dynamics. 

The $J/\psi$ production in this study is based on the freeze-out
phase space information of charm quarks from the AMPT model.
The AMPT model is a transport model that simulates relativistic
heavy ion collisions \cite{Zhang:1999bd,Lin:2000cx,%
Lin:2001yd,Lin:2004en}. 
It uses the HIJING model \cite{Wang:1991ht} 
to provide initial conditions.
Either mini-jet partons, or partons from string melting will
participate in the space-time evolution of the system.
The parton evolution is carried out by the ZPC parton 
cascade model \cite{Zhang:1997ej}.
At parton freeze-out, partons are converted into hadrons using
either the Lund string fragmentation model
or a coordinate-space coalescence
model that combines nearest partons into hadrons. Then the ART
hadronic transport model \cite{Li:1995pr,Li:2001xh} 
is used to evolve the hadronic system.
The AMPT model can reasonably describe particle distributions
at RHIC. It is also successful in showing the importance of
partonic evolution on elliptic flow and HBT 
radii \cite{Lin:2001zk,Chen:2004dv,Lin:2002gc}. 
In Ref.~\cite{Zhang:2005ni}, charm flow in Au+Au collisions
at $\sqrt{s_{NN}}=200$ GeV is studied using
the AMPT model in the string melting scenario
with the perturbative method. The initial $D$ mesons
follow a parametrization of $D$ meson $p_t$ distributions from the 
STAR collaboration \cite{Tai:2004bf} 
and a rapidity plateau between -2 and +2. 
They are dissociated into charm quarks. Screened Coulomb
cross sections are used for the rescatterings. Results from
the 3mb cross section case are compared to those using a 10mb
cross section. The $D$ meson elliptic flow and non-photonic electron
elliptic flow are found to be very sensitive to the 
charm rescattering cross sections.

In the following, the charm quark freeze-out information
will be used for the study of
production of the $J/\psi$ particle using a
phase-space coalescence 
model \cite{Mattiello:1995xg,Mattiello:1996gq,Chen:2003qj,%
Chen:2003av}. 
In this model, the $J/\psi$ momentum distribution is given by
\begin{equation}
\frac{d^3N_{J/\psi}}{d^3p}=g_{J/\psi}
\int \frac{d^3q\;d^3R\;d^3r}{(2\pi)^{3\times 2}}
f_c(\vec{x}_1,\vec{p}_1)f_{\bar{c}}(\vec{x}_2,\vec{p}_2)
f^W_{J/\psi}(\vec{r},\vec{q}).
\end{equation}
In the above formula, $g_{J/\psi}=1/12$ is the degeneracy factor
of $J/\psi$ production from charm and anti-charm quarks. $\vec{q}$ 
is the relative momentum, $\vec{r}$ is the relative position, 
and $\vec{R}$ is the center-of-mass position of a pair. $f_c$,
$f_{\bar{c}}$ are the freeze-out phase-space distributions of charm
and anti-charm quarks. $f^W_{J/\psi}$ is the $J/\psi$ Wigner
function. With a Gaussian spatial wave function,
\begin{equation}
f^W_{J/\psi}(\vec{r},\vec{q})=
8\exp\left(-\frac{r^2}{\sigma^2}-q^2\sigma^2 \right).
\end{equation}
The width $\sigma$ is related to the $J/\psi$ rms radius by
$r^2_{rms}=\frac{3}{8}\sigma^2$. 
$r_{rms}$ is taken to be 0.5 fm as given by the potential
model \cite{Eichten:1979ms}.
$J/\psi$ production from two freeze-out distributions will
be compared. One has 3mb parton rescattering cross sections,
and the other has 10mb cross sections. 
Only $J/\psi$ production from coalescence is taken into
account. There is no production and survival from initial
nucleon-nucleon collisions. The above coalescence
approach can not accurately account for binding energy.
No feeddowns from higher charmonium resonances are taken
into account.

\begin{figure}[hbt]
 \includegraphics[scale=0.8]{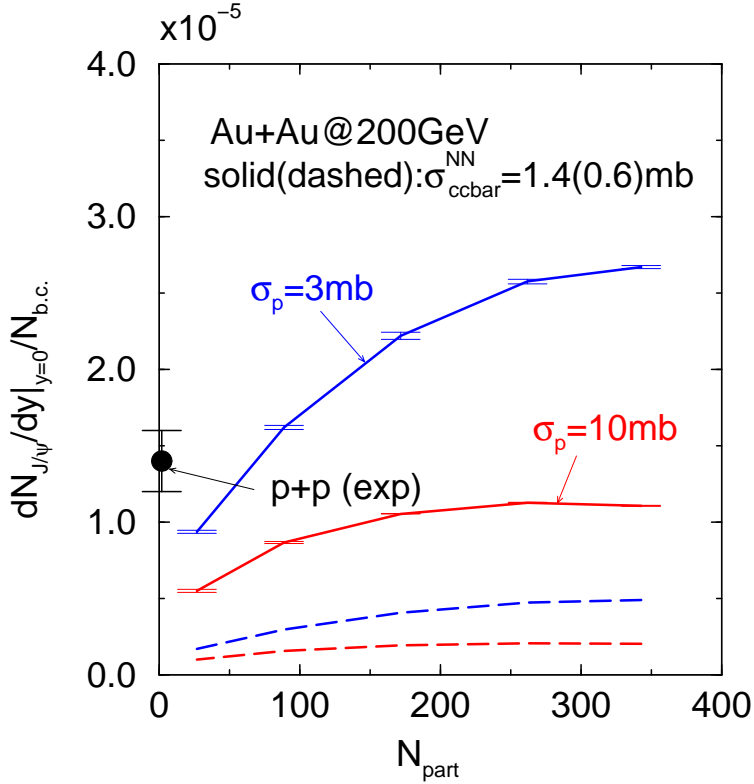}
 \caption{\label{fig_jpsi_dndy_nc_cent}$J/\psi$ yield per binary
nucleon-nucleon collision as a function of the number of 
participant nucleons. The data point is from Ref.~\cite{Adler:2005ph}.}
\end{figure}

The $J/\psi$ rapidity density per binary nucleon-nucleon collision
as a function of the number of participant nucleons
is shown in Fig.~\ref{fig_jpsi_dndy_nc_cent}. As expected, the
$J/\psi$ yield increases with the number of participants. The
3mb case has a larger yield compared to the 10mb case. In central
collisions, it can be larger by a factor of about 2.4. This
is because of charm and anti-charm quarks are closer in phase-space
with smaller cross section and less rescatterings. The solid 
curves are for the case with 1.4mb for the charm production cross section 
in nucleon-nucleon collisions \cite{Adams:2004fc}.
At the moment, the charm production cross section has large uncertainty.
When the cross section goes down to 0.6mb \cite{Adler:2004ta}, 
the $J/\psi$ yield decreases
to about 20\% of the 1.4 mb result. For central collisions, 
both the 3mb case and the 10mb
case can give suppression consistent with preliminary PHENIX
data \cite{PereiraDaCosta:2005xz} 
relative to the production in $p+p$ collisions \cite{Adler:2005ph}.
It is also interesting to see that the centrality dependence
has almost the same shape as the production from recombination,
e.g., from Grandchamp and Rapp's 
calculations \cite{Grandchamp:2003uw}. In other words,
coalescence and recombination are closely related to each other.
If the survival of charmonium from initial nucleon-nucleon
collisions is taken into account, the $J/\psi$ results are expected to
be similar to those of Grandchamp and Rapp.

\begin{figure}[hbt]
 \includegraphics[scale=0.8]{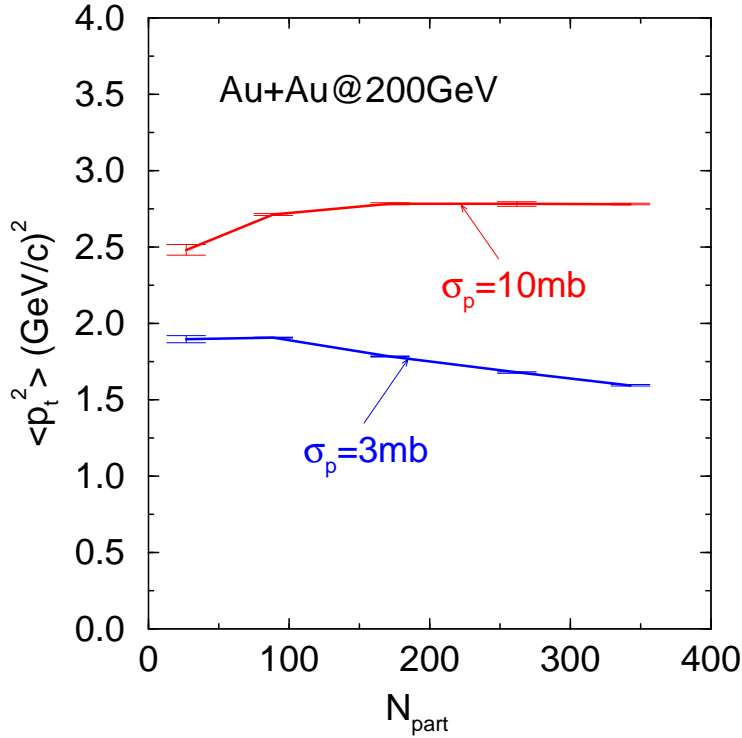}
 \caption{\label{fig_jpsi_avgpt1}Averaged $p_t^2$ at mid-rapidity
as a function of the number of participant nucleons.}
\end{figure}

The averaged $p^2_t$ as a function of the number of participants
is shown in Fig.~\ref{fig_jpsi_avgpt1}. The 10 mb case has
more radial flow and is above the 3mb case. The 
$\langle p^2_t \rangle$ increases with centrality as
more radial flow is generated in the 10 mb case. The 3mb
case is slightly different. A closer look at the $p_t$ distributions
reveals that in peripheral collisions, high $p_t$ charm quarks
escape easily and are not affected, while in central
collisions, the quenching of high $p_t$ charm leads
to a decrease of $\langle p^2_t \rangle$ of $J/\psi$ particles.
The $\langle p^2_t \rangle$ is about $3$ GeV$^2$ in the
10 mb case and is about 1.5 GeV$^2$ in the 3 mb. The
10 mb case is comparable to recent calculations from recombinations
by Thews \textit{et al.} \cite{Thews:2005vj,Thews:2006mm} 
and also comparable to preliminary
PHENIX central electron arm data \cite{PereiraDaCosta:2005xz}. 

\begin{figure}[hbt]
 \includegraphics[scale=0.8]{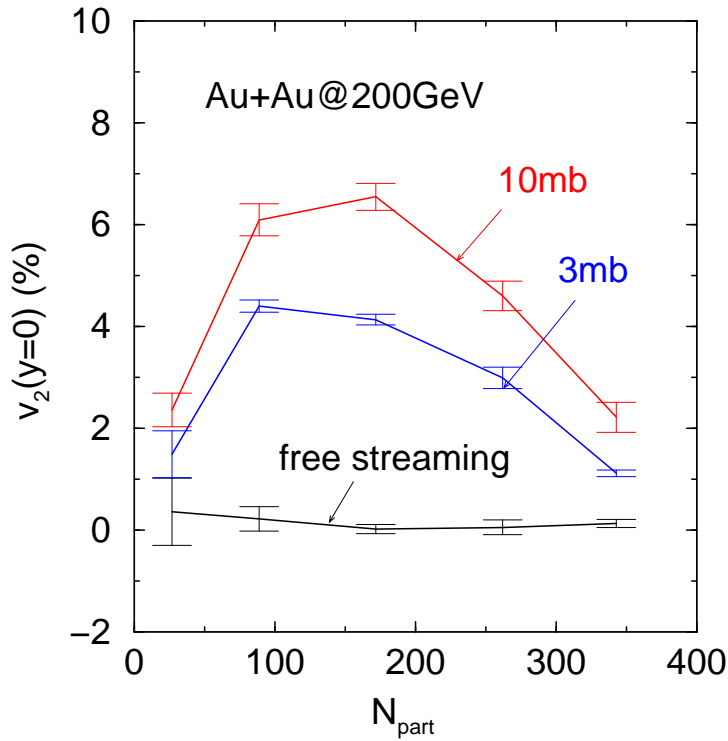}
 \caption{\label{fig_jpsi_v2y_cent1}Elliptic flow as a function
of the number of participant nucleons.}
\end{figure}

Fig.~\ref{fig_jpsi_v2y_cent1} shows the elliptic flow parameter
$v_2$ as a function of centrality. The shapes of the curves
are similar to those of charged hadrons. However, different
from the charged hadron case \cite{Lin:2004en}, 
the elliptic flow is more 
sensitive to the cross section. 
This is due to the increased sensitivity of radial flow of
massive particles relative to light hadrons as seen in
Fig.~\ref{fig_jpsi_avgpt1}. Larger cross section has
larger asymptotic elliptic flow at high $p_t$, at the
same time it has larger $\langle p_t^2 \rangle$ which
leads to more weight of high $p_t$ flow in the integrated $v_2$.

\begin{figure*}[hbt]
 \includegraphics[scale=0.8]{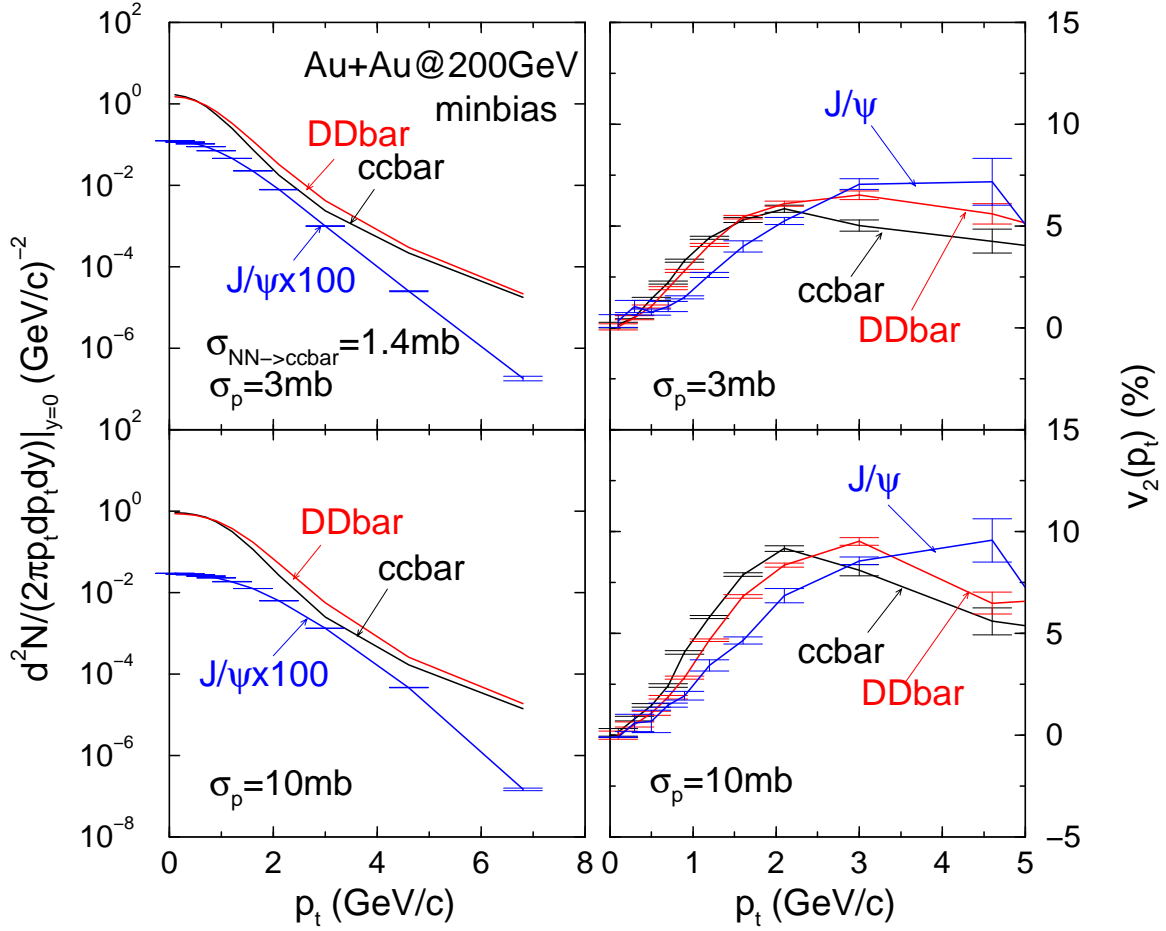}
 \caption{\label{fig_cDJ_dndpt_v2pt1}Invariant transverse momentum
distributions and $p_t$ differential elliptic flows for 
charm quarks, $D$ mesons, and $J/\psi$ particles.}
\end{figure*}

More details can be seen by looking at the $p_t$ distributions
and the $p_t$ differential $v_2$ curves. They are shown
in Fig.~\ref{fig_cDJ_dndpt_v2pt1} for minimum-bias
Au+Au collisions at $\sqrt{s_{NN}}=200$ GeV. 
In addition to the $J/\psi$ results, the charm quark and
the $D$ meson (including $D^*$ meson) results are also shown.
Being affected by both the charm and the light quarks, 
$D$ mesons have an invariant $p_t$ distribution that has the same
concave shape as the charm quark distribution with a slightly higher
averaged $p_t$. On the other hand, $J/\psi$ comes only from
charm quarks. The $p_t$ distribution has a different, convex,
shape, with a more enhanced averaged $p_t$.

The elliptic flow results show mass ordering for 
the low $p_t$ region. Charm and $D$ meson
$v_2$ curves increase with $p_t$, reach a peak, then decrease a little.
$J/\psi$ $v_2$ in the $p_t$ range shown here increases, crosses those
of charm quarks and $D$ mesons up to a value that is comparable
to the peak value of charm and $D$ meson elliptic flow. This
behavior is different from some previous studies in 
which $J/\psi$ $v_2$ is consistently larger than
$D$ meson $v_2$ \cite{Lin:2003jy}. The crossing of
the $J/\psi$ and $D$ meson $v_2$ curves reflects the
distinct freeze-out phase space distributions from the
AMPT model.

In summary, $J/\psi$ production from the coalescence of charm
and anti-charm quarks reflects charm interactions in the QGP. The
$J/\psi$ yield is very sensitive to the charm production cross
section in $p+p$ collisions. Coalescence yield has a centrality
dependence that is similar to that of recombination models. 
Both the 3mb and 10mb cases can lead to suppression on the
same level as recombination models and consistent with
preliminary PHENIX results. The 10mb case gives 
$\langle p^2_t \rangle$ around 3 GeV$^2$ and is comparable
to recombination results and preliminary PHENIX data, while
the 3mb results are much lower than the 10mb case. The elliptic
flow follows the same centrality dependence as charged hadrons,
but is more sensitive to the cross section partly due to
the enhanced sensitivity of $p_t$ distribution because of
the large mass of the $J/\psi$ particle. The $J/\psi$ $p_t$
distribution is convex and different from that of $D$ mesons
because of the combination of charm and anti-charm quarks. The
$J/\psi$ elliptic flow crosses that of $D$ mesons and reaches about
the peak value of the $D$ meson elliptic flow. Hence, the combination
of open and hidden charm measurements will provide a
more complete picture of the evolution of the charm quarks
in the strongly interacting Quark-Gluon Plasma.

\begin{acknowledgments}

Discussions with A. Baltz, L.W. Chen, G. Fai, C.M. Ko, L. McLerran,
E. Mottola are greatly appreciated. B.Z. thanks the Parallel Distributed
Systems Facilities for providing computing resources. This work is 
supported by the U.S. National Science Foundation under
grant No's PHY-0140046 and PHY-0554930.
\end{acknowledgments}

\end{document}